                                                                              
\magnification=\magstep1

\hoffset=-0.3truecm 

\hsize = 33pc
\vsize = 46.5pc
\baselineskip=14pt
\tolerance 2000
\parskip=4pt

\font\grand=cmbx10 at 14.4truept
\font\grand=cmbx10 at 14.4truept

\font \ninerm                 = cmr9

\pageno=0
\def\folio{
\ifnum\pageno<1 \footline{\hfil} \else\number\pageno \fi}

\def\q#1{[#1]}              
\def\bibitem#1{\parindent=8mm\item{\hbox to 6 mm{\q{#1}\hfill}}}
\font\extra=cmss10 scaled \magstep0
\setbox1 = \hbox{{{\extra R}}}
\setbox2 = \hbox{{{\extra I}}}
\setbox3 = \hbox{{{\extra C}}}

\def\C{{{\extra C}}\hskip-\wd3\hskip2.5 true pt{{\extra I}}\hskip-\wd2
\hskip-2.5 true pt\hskip\wd3}
\def\Complex{\hbox{{\extra\C}}}   
\setbox4=\hbox{{{\extra Z}}}

\phantom{nothing}

\vskip 2.0truecm

\centerline
{\grand  Extended matrix Gelfand-Dickey hierarchies: }

\centerline{\grand reduction to classical Lie algebras}

\vskip 1.0truecm

\centerline{
L\'aszl\'o Feh\'er${}^{a}$ \  and \  Ian Marshall${}^{b}$}

\bigskip

\centerline{\it 
${}^{a}$Physikalisches Institut der Universit\" at Bonn,
Nussallee 12,  53115 Bonn, Germany}
\centerline{\it e-mail: feher@avzw01.physik.uni-bonn.de}

\bigskip

\centerline{\it ${}^{b}$Department of Mathematics, Leeds University, 
Leeds LS2 9JT, U.K.}
\centerline{\it e-mail: amt6im@amsta.leeds.ac.uk}

\vskip 1.0 truecm 
\centerline{\bf Abstract}
\medskip

The Drinfeld-Sokolov reduction method  has been used to associate 
with $gl_n$ extensions of the matrix $r$-KdV system. 
Reductions of these systems to the 
fixed point sets of involutive Poisson  maps, implementing reduction of 
$gl_n$ to classical Lie algebras of type $B$, $C$, $D$, are here presented.
Modifications corresponding, in the first place to factorisation of the 
Lax operator, and then to Wakimoto realisations of the current algebra 
components of the factorisation, are also described. 
                                                                               
\vfill\eject

\def \pa {\partial}	\def \paw {(\pa+w)^{-1}}	
\def \cinf {C^\infty}	\def\Tr {{\rm{Tr}}\,} 	\def\tr {{\rm{tr}}\,}
\def \and {{\hbox{and }}}	\def\mat{{\rm{mat}}}
\def \MDS {{\cal M}_{\rm DS}} \def\res{{\rm res}}
\def\vt{\vartheta}  \def\G{{\cal G}}
\def\paw{(\pa+w)^{-1}}	\def\pad{(\pa+d)^{-1}}
\def\dt {\left.{d\over dt}\right\vert_{t=0}}
\def\symp{{\cal S}}

\def\Ch{{1}}
\def\OS{{2}}
\def\Oe{{3}}
\def\De{{4}}
\def\Liu{{5}}
\def\Che{{6}}
\def\Di{{7}}
\def\Arat{{8}}
\def\FM{{9}}
\def\DS{{10}}
\def\KP{{11}}
\def\GHM{{12}}
\def\McI{{13}}
\def\F{{14}}
\def\KW{{15}}
\def\FG{{16}}
\def\dBF{{17}}
\def\A{{18}}
\def\D{{19}}
\def\DF{{20}}
\def\Car{{21}}

\centerline{\bf 1.~Introduction}
\medskip

We consider examples of constrained KP hierarchies having a 
Lax operator of the form
$$
L = \ell + z_+\paw z_-
\qquad\hbox{with}\qquad
\ell= \Delta^r\pa^r + u_1\pa^{r-1}+ \cdots + u_r ,
\eqno(1)$$
where $\Delta$ is a constant diagonal matrix with $\Delta^r$ having 
distinct, non-zero entries, $u_0,\dots,u_r$ are in $\widetilde{gl}_p$, 
$z_+\in\widetilde{\mat}(p\times s)$, $z_-\in\widetilde{\mat}(s\times p)$ 
and $w$ is in $\widetilde{gl}_s$.
Here $\mat(m\times n)$ denotes the set of $m\times n$ complex 
matrices and for any
vector space $V$, $\tilde V$ stands for $\cinf(S^1,V)$. $PDO(m\times n)$
denotes the set of pseudodifferential operators with coefficients in
$\widetilde{\mat}(m\times n)$.
We shall make use of the standard splitting
$PDO=PDO_++PDO_-$ of the space of pseudodifferential operators as a
vector space direct sum of differential operators and integration
operators. We also use the standard trace-form ``res'' on $PDO$ given by
$\res\sum a_i\pa^i=a_{-1}$.

We call the systems associated with Lax operators of the form given in (1)
{\it systems of extended Gelfand-Dickey type}.
These systems are defined for any integers 
$r,p\geq 1$ and $s\geq 0$.
For simplicity of language, we shall formulate our statements 
having in mind the generic case for which $r>1$ and $s>0$.
Note however that all statements are also valid in the special cases 
for which either $r=1$ or $s=0$, even though some of them become trivial.
The special cases  for which $r>1$ and $s=0$ reproduce the standard  
$p\times p$  matrix  Gelfand-Dickey systems.
The cases  with  $r=1$  correspond to generalised AKNS systems.

There have been several papers over the
last few years devoted to systems of the above type \q{\Ch--\Arat}.
It was shown in \q{\FM} how hierarchies  with a Lax operator 
of the form
in (1) can be obtained by the Drinfeld-Sokolov (DS) reduction method.
Specifically, with the partition 
$$
n=pr+s=\overbrace{r+\cdots +r}^{p\;\rm times}+
\overbrace{1+\cdots +1}^{s\;\rm times}
\eqno(2)
$$
is associated \q{\KP} a graded  Heisenberg subalgebra of the 
loop algebra $gl_n\otimes \Complex[\lambda,\lambda^{-1}]$, 
and a generalised KdV hierarchy having the Lax operator in (1) results 
from application of the DS reduction procedure \q{\DS}
(see also \q{\GHM,\McI,\F}) with respect to a 
grade-one element from this Heisenberg subalgebra, if $r>1$. 
In the $r=1$ special case the DS reduction becomes trivial, 
but interesting results remain valid.

The Lax operator usually studied in the literature for $s>0$ is 
obtained from (1) by choosing $p=1$ and $w=0$. 
In fact setting $w=0$  is not 
advantageous since this Dirac reduction of 
the phase space leads to non-local Poisson brackets.

In the present paper we investigate the discrete
symmetries given by involutive Poisson maps on the phase space 
of an extended Gelfand-Dickey system. 
Reduction to the fixed point set of such a map yields 
systems which arise from using the classical Lie algebras $B$, $C$, $D$ in 
the DS approach.

The further purpose of the paper is to study modifications of the 
above systems. 
In principle, modification arises
via two possible mechanisms. The first can be viewed as an
application of the well-known factorisation approach of 
Kupershmidt-Wilson \q{\KW} (see also \q{\FG}). 
The phase space of the resulting modified system is a direct 
product whose factors carry linear Poisson structures,
typically given by current algebras. 
The second is a novel construction which involves the 
so-called Wakimoto realisations of the  current algebras, 
as was described in \q{\dBF}. 
We will show that the two mechanisms are in fact closely related.

We shall use the abbreviation  ``PB'' for
Poisson bracket and  shall refer to the first and second 
Adler-Gelfand-Dickey PBs  on $PDO(p\times p)$ \q{\A,\D} as the ``AGD PBs''.

\bigskip
\centerline{\bf 2.~Extended Gelfand-Dickey hierarchies}
\medskip

In this section we list the main elements of the theory of
systems of extended Gelfand-Dickey type. 
Many of the results are described fully in \q{\FM}, whilst at 
the same time 
much of this theory is standard and goes back to the work of 
 Adler, Gelfand-Dickey and Drinfeld-Sokolov,
see \q{\A}, \q{\D} and \q{\DS}.

Let $\MDS$\footnote{${}^1$}{\ninerm In the 
Hamiltonian reduction approach \q{\FM} $\MDS$ represented a 
so-called DS gauge: we have kept this nomenclature here.}
be the space of quadruples
$(\ell, z_+, z_-, w)$ that appear in  (1), i.e., as a space 
$$
\MDS=\bigl(\widetilde{gl}_p\bigr)^r
\times\widetilde{\mat}(p\times s)
\times\widetilde{\mat}(s\times p)\times \widetilde{gl}_s .
\eqno(3)
$$
The functions on $\MDS$  of interest are local functionals 
that have the form 
$H=\int_{S^1} h(u_1,\dots,u_r,z_+,z_-,w)$ 
with $h$ a differential polynomial in the entries of the matrices in 
its arguments.
There are two compatible PBs on $\MDS$, 
given by the following formulae for the respective hamiltonian 
vector fields: 
$$
{\bf X}^1_H(\ell)=
	\left[\ell, {\delta H\over \delta \ell}\right]_+, \quad
{\bf X}^1_H(z_\pm )=\pm  {\delta H\over \delta z_\mp},\quad
{\bf X}^1_H(w)=0,
\eqno(4)$$
and 
$$
\eqalign{
&{\bf X}_H^2(\ell)
=
\left(\ell {\delta H\over \delta \ell}\right)_+ \ell
-\ell\, \left({\delta H\over \delta \ell} \ell\right)_+
+ \left(\ell{\delta H\over\delta z_-}\paw z_-\right)_+
-\left(z_+\paw{\delta H\over \delta z_+}\ell\right)_+ \cr
&{\bf X}_H^2(z_+)=\res\left(\ell\left({\delta H\over \delta\ell}z_+
+{\delta H\over \delta z_-}\right)\paw\right) - z_+{\delta H\over \delta w}
\cr
&{\bf X}_H^2(z_-)=-\res\left(\paw\left(z_-{\delta H\over \delta\ell}
+{\delta H\over \delta z_+}\right)\ell\right)+
{\delta H\over \delta w}z_-\cr
&{\bf X}_H^2(w)={\delta H\over \delta z_+}z_+ - 
z_-{\delta H\over \delta z_-}
+\left[ {\delta H\over \delta w}, w\right] -
\left({\delta H\over \delta w}\right)'.\cr}
\eqno(5)
$$
Here the gradients are defined by
$$
\dt H(\ell +t\delta\ell, 
z_\pm+t\delta z_\pm, w+t\delta w)
=\Tr\left({\delta H\over\delta\ell} \delta\ell \right) +
\int_{S^1}\tr \left( 
{\delta H\over\delta z_+} \delta z_+ +{\delta H\over\delta z_-} \delta z_-
	+ {\delta H\over \delta w} \delta w \right),
\eqno(6)
$$
where $\Tr$ stands for $\int\tr \res$ and 
$
{\delta H\over\delta\ell}=\sum_{i=1}^r\pa^{i-r-1}{\delta H\over\delta u_i}.
$
For any $A\in PDO$,  we have the decomposition 
$A=A_+ + A_-$ defined by the standard splitting of $PDO$.

The map $\pi: \MDS \to PDO(p\times p)$, which assigns the
pseudodifferential operator  $L$ in
(1) to the point $(\ell, z_+,z_-, w)$ in $\MDS$, 
is a Poisson map with respect to the PBs defined by the formulae (4) and
(5) on $\MDS$ and the first and second  AGD  PBs on
$PDO(p\times p)$, respectively.
It follows that $M=\pi(\MDS)$ --- the set of Lax operators 
of the form  (1) --- is a Poisson subspace of $PDO(p\times p)$ with 
respect to the first and second AGD PBs. 

To define the commuting flows on $\MDS$ we proceed as follows.
We first diagonalise the Lax operator $L=\pi(\ell, z_+,z_-, w)$.
That is to say we write
$$
L = g\hat L g^{-1}
\eqno(7)$$
for $g$ an element of $PDO(p\times p)$ of the form
$g = {\bf 1}_p + \sum_{k=1}^\infty g_k\pa^{-k}$
and we require that $\hat L$  be diagonal and $g$ be off-diagonal,
which determines them uniquely.
For $Q$ a constant, diagonal $p\times p$ matrix,
let the functions $H_j^Q$ be defined by
$$
H_0^Q(\ell,z_\pm,w) = \Tr\left(\hat LQ (\Delta \pa)^{-r}\right),
\quad
H_j^Q(\ell,z_\pm,w) ={r\over j} \Tr\left(\hat L^{j/r}Q\right)
\quad{\hbox{for }}j=1,2,\dots\,.
\eqno(8)
$$
The set of functions $H_j^Q$ for  $j=0,1,2,\dots$ and $Q$ arbitrary 
yields commuting Hamiltonians on $\MDS$. 
The corresponding hamiltonian vector fields   are conveniently
expressed in the following form:
$$\eqalign{
{\bf X}^2_{j,Q}(L)
&= {\bf X}^1_{j+r, Q}(L)\ =
\left[ \left( g Q g^{-1} L^{j/ r}\right)_+ , L\right]\cr
{\bf X}^2_{j,Q}(z_+)
&= {\bf X}^1_{j+r, Q}(z_+)=
\res\left( g Q g^{-1} L^{j/ r} z_+ \paw\right)\cr
{\bf X}^2_{j,Q}(z_-)
&= {\bf X}^1_{j+r, Q}(z_-)=
-\res\left( \paw z_-g Q g^{-1} L^{j/r} \right)\cr
{\bf X}^2_{j,Q}(w)
&= {\bf X}^1_{j+r, Q}(w)\ = 0,\qquad\qquad \forall j=0,1,\ldots\,.\cr}
\eqno(9)
$$
These commuting vector fields 
generate the flows of  the extended Gelfand-Dickey hierarchy.
If $r=p=1$ and $s=0$, then  the flows are trivial, and we henceforth exclude 
this case.

\bigskip
\centerline{\bf 3.~Modifications of extended Gelfand-Dickey hierarchies}
\medskip

We next apply a two-step factorisation procedure to the 
Lax operator $L$ that leads to modifications of the flows in (9).
By {\it modification}, we mean that there is a non-invertible 
Poisson map given in terms of a differential polynomial formula,
from the Poisson space of the new (modified) variables to $\MDS$.
The Hamiltonians of the modified flows are the pull-backs of the functions
$H_j^Q$ in (8).
The first step of the factorisation procedure is rather well-known 
\q{\De,\Liu,\Che,\FM}.
The second step was mentioned in passing in \q{\FM} but details were
omitted. Here we also explain the relationship of this second step to
the Wakimoto realisations of the current algebra based on the
general linear Lie algebra. 

Let us introduce the space 
$\Theta=\bigl(\widetilde{gl}_p\bigr)^{r-1}\times  \widetilde{gl}_{p+s}$
and endow it with the current algebra PB on each of the components.
The points of this space are denoted as 
$(\theta_1,\dots,\theta_{r-1}, \theta_r)\in\Theta$.
For local functionals $F,H$ on $\Theta$ we thus have 
$$
\{F,H\}(\theta_1,\dots,\theta_{r-1},\theta_r)=
\sum_{i=1}^{r}\int\tr\left(\theta_i
\left[{\delta F\over\delta\theta_i},{\delta H\over\delta\theta_i}\right]-
{\delta F\over\delta\theta_i}
\left({\delta H\over\delta\theta_i}\right)'\right).
\eqno(10)$$
There is a Poisson map $\mu$  from $\Theta$ to $\MDS$ 
described in \q{\FM}.
It is important to note that $\mu$ is Poisson with respect to 
the {\it second} Poisson structure on $\MDS$ given by (5), and {\it not}
with respect to the first Poisson structure given by (4).
We shall not 
specify $\mu$ here, but we give the form of the composition 
$\Phi = \pi \circ \mu :\Theta\rightarrow M\subset PDO(p\times p)$.

Let us write the matrix $\theta_r\in\widetilde{gl}_{p+s}$ in the form 
$$
\theta_r=\left(\matrix{a&b\cr c&d}\right),
\eqno(11)
$$
where $a\in\widetilde{gl}_p$, $b\in\widetilde{\mat}(p\times s)$, 
$c\in\widetilde{\mat}(s\times p)$,  $d\in\widetilde{gl}_s$.
Fix an integer $\kappa$ between $0$ and $r-1$.
Then $\Phi$ is given by
$$
L=\Delta(\pa+\theta_1)\Delta (\pa+\theta_2)\cdots
\Delta(\pa+\theta_\kappa)\Delta
\bigl[\pa+a-b\pad c\bigr]
\Delta(\pa+\theta_{\kappa+1})\cdots\Delta(\pa+\theta_{r-1}).
\eqno(12)$$
For $\kappa=0$ there are no factors of the form $(\pa+\theta_i)$ on the left, 
while for $\kappa=r-1$ there are none on the right.
Of course different choices 
of $\kappa$ correspond to different definitions of $\mu$ but all of them are 
related by invertible transformations. 
Hence all of the apparently different modifications for the different
choices of $\kappa$ are equivalent.

As the composition of two Poisson maps, $\Phi$ is guaranteed to 
be a Poisson map
with respect to the second AGD PB on $PDO(p\times p)$.
A direct proof of the Poisson property of $\Phi$ 
can be obtained using the following results.

\noindent
{\bf Lemma 1:}\quad 
{\it The multiplication map  $: PDO\times PDO\rightarrow PDO$
defines a Poisson map with respect to  the second AGD PB on $PDO$.}

\noindent
{\bf Lemma 2:}\quad
{\it $\{\pa + \theta\,\vert\,\theta\in\widetilde{gl}_p\}\subset 
PDO(p\times p)$
is a Poisson subspace with respect to  the second AGD PB,
which on this subspace coincides with the current algebra PB appearing 
in (10) for $i\neq r$.}

\noindent
{\bf Lemma 3:}\quad
{\it The map
${\eta}:\widetilde{gl}_{p+s}\rightarrow PDO(p\times p)$ defined by
$$
\eta\left(\matrix{a&b\cr c&d}\right) = 
\pa + a - b\pad c
\eqno(13)
$$
is a Poisson map with respect to the current algebra PB on 
$\widetilde{gl}_{p+s}$ that occurs in (10) for $i=r$ 
and the second AGD PB on $PDO(p\times p)$.}

The first two lemmas seem to be part of the general knowledge in
the field of integrable hierarchies. 
Lemma 3 was proved in \q{\FM}.
The modified flows are defined on the phase space $\Theta$ by pulling back
the Hamiltonians $H^Q_j$ in (8) by means of the map $\mu$.

For reasons explained in \q{\FM} (see also \q{\Ch,\OS}), 
if  $s\neq 0$ the factor 
$$
K:= \pa + a - b\pad c 
\eqno(14)$$
entering the factorisation of $L$ in (12) is called the ``AKNS factor''.
Then the flows on $\Theta$ can themselves be modified by 
factorising $K$ as follows 
$$
K = \pa + a - b\pad c =
(\pa+\vartheta_0)
\left({\bf 1}_p -\gamma(\pa+\vartheta_1+\beta\gamma)^{-1}\beta\right)
\eqno(15)$$
for
$$
(\vartheta_0, \vartheta_1, \beta, \gamma)\in 
\widetilde{gl}_p\times\widetilde{gl}_s\times \symp
\quad\hbox{where}\quad 
\symp:=\widetilde{\mat}(s\times p)\times\widetilde{\mat}(p\times s).
\eqno(16)$$
We let the space 
$\widetilde{gl}_p\times\widetilde{gl}_s\times\symp $ be endowed with
the natural direct sum Poisson structure. That is if $F$, $H$ are two
local functionals on this space, we have
$$\eqalign{
\{F&,H\}(\vt_0, \vt_1, \beta,\gamma)=\cr
&\sum_{i=0,1} \int_{S^1}\tr\left(
\vt_i \left[{\delta F \over \delta \vt_i},
{\delta  H \over \delta \vt_i}\right]
-{\delta  F  \over \delta \vt_i}
\left({\delta H \over \delta\vt_i}\right)'\right)    
+\int_{S^1}\, {\rm tr}\left(
{\delta  F \over \delta \beta} {\delta H \over\delta\gamma}
-{\delta  H \over \delta \beta} {\delta F \over \delta\gamma}\right).
\cr}\eqno(17)
$$
The factorisation specified in (15) can be lifted to a mapping 
$\nu: \widetilde{gl}_p\times\widetilde{gl}_s\times \symp
 \rightarrow \widetilde{gl}_{p+s}$ whose equation is
$$
a=\vartheta_{0}-\gamma\beta,\quad
d=\vartheta_1+\beta \gamma,\quad
b=\vartheta_{0}\gamma -\gamma\vartheta_1 - \gamma\beta \gamma+\gamma',
\quad c=\beta
\eqno(18)
$$
and direct calculation proves the following 

\noindent
{\bf Proposition 4:}
{\it If 
$\widetilde{gl}_p\times\widetilde{gl}_s\times\symp $ 
and $\widetilde{gl}_{p+s}$ 
are endowed with the PB in (17)
and with the current algebra PB, respectively, then 
the map $\nu$ determined by (18) is a Poisson map.} 

Define the space 
$\Theta'= \bigl(\widetilde{gl}_p\bigr)^{r-1}\times 
\widetilde{gl}_p\times\widetilde{gl}_s\times\symp$ and endow it with
the product PB given by
the current algebra PB on  $\bigl(\widetilde{gl}_p\bigr)^{r-1}$
together with the PB in (17).
Then $\nu$ gives rise to a Poisson map $\nu': \Theta' 
\rightarrow \Theta$,
which acts as $\nu$ on 
$\widetilde{gl}_p\times\widetilde{gl}_s\times \symp$
and as the identity on the  
$\bigl(\widetilde{gl}_p\bigr)^{r-1}$ factor.
This map provides us with a modification of the system on 
$\Theta=\bigl(\widetilde{gl}_p\bigr)^{r-1}\times \widetilde{gl}_{p+s}$.
The resulting modified system is the same as the one engendered 
by the composite Poisson map $\mu\circ \nu': \Theta' \rightarrow \MDS$.

We now explain that the map $\nu$ defined by (18) can be used 
to generate  a huge family of  ``realisations''  of the current algebra 
PB based on $gl_m$  for any $m$. 
For this we simply repeat the construction for an arbitrary partition of 
$m$ of the form $m=m_1+m_2$. 
This amounts to writing $\theta\in \widetilde{gl}_m$ as 
$\theta=\left(\matrix{a&b\cr c&d}\right)$ with $a\in \widetilde{gl}_{m_1}$
etc, and expressing $a,b,c,d$ by
formula (18) in which we then insert the variables 
$$
(\vartheta_0, \vartheta_1, \beta, \gamma)\in 
\widetilde{gl}_{m_1}\times\widetilde{gl}_{m_2}\times \symp_{m_1,m_2}
\quad\hbox{with}\quad 
\symp_{m_1,m_2}:=
\widetilde{\mat}(m_2\times m_1)\times
\widetilde{\mat}(m_1\times m_2).
\eqno(19)$$
The PBs of these new variables defined similarly to (17) then imply the
current algebra PB for the variable $\theta$, that is we have a 
Poisson map like in Proposition 4.
Repeating the construction iteratively for the current algebra factors,
we can associate a Poisson map 
$$
\nu_{m_1, m_2,\ldots, m_l}:
\widetilde{gl}_{m_1}\times \widetilde{gl}_{m_2}\times \cdots \times 
\widetilde{gl}_{m_l} \times \symp_{m_1,m_2,\ldots, m_l}
\rightarrow 
\widetilde{gl}_{m}
\eqno(20)$$
with any partition $m=m_1 + m_2 +\cdots +m_l$.
The procedure gives that as a vector space 
$$
\symp_{m_1,m_2,\ldots, m_l}=\widetilde{\Complex}^d\times 
\widetilde{\Complex}^d
\quad\hbox{for}\quad 
2d=m^2 - m_1^2 -m_2^2 - \cdots - m_l^2,
\eqno(21)$$
and it carries the corresponding canonical PB.
The precise formula of the map in (20) depends 
on the route  whereby the final partition of $m$ is reached
through the iterative procedure.
However, it has been proved in \q{\dBF} (in a more general context) that
the various Poisson maps that follow are all related by invertible 
Poisson maps. The map in (20) is known as a  generalised (classical) 
Wakimoto realisation of the current algebra PB based on $gl_m$.
The standard  Wakimoto realisation 
belongs to the partition of $m$ for which all $m_i=1$.
An explicit  formula for the  Wakimoto realisations 
was derived in \q{\dBF} by different methods.
Further results  and background on Wakimoto realisations 
can also be  found in \q{\dBF} and references therein.

We can use any of the  Wakimoto 
realisations in (20) to modify any of the current algebra factors that 
appear in the factorisation of $L$ in (12).
This yields a large family of modifications of the extended 
Gelfand-Dickey hierarchies.

\bigskip
\centerline{\bf 4.~Discrete reductions}
\medskip

We now search for discrete symmetries of the 
extended Gelfand-Dickey systems.
The compatible PBs on the phase space $\MDS=\{(\ell, z_+, z_-, w)\}$ 
are given by   
$$\eqalign{
\{ F, H\}^*_i&= {\bf X}_H^i(F) =\cr
&{\rm Tr}\left({\delta F\over \delta \ell}{\bf X}_H^i( \ell)\right)
+
\int_{S^1} {\rm tr}\left(
{\delta F\over \delta z_+}{\bf X}_H^i( z_+)
+{\delta F\over \delta z_-}{\bf X}_H^i(z_-)
+{\delta F\over \delta w}{\bf X}_H^i(w)\right)\cr}
\eqno(22)$$ 
for arbitrary local functionals  $F, H$ on $\MDS$,
where ${\bf X}_H^i$  are defined by  (4), (5) for $i=1, 2$.
Specifically, we look for  symmetries given
by some  involutive map 
$$
\sigma : \MDS \rightarrow \MDS,
\qquad
\sigma^2 = {\rm id},
\eqno(23)$$
which  leaves the PBs invariant,
$$
\{ F\circ \sigma , H\circ \sigma\}_i^*=\{ F, H\}_i^*\circ \sigma,
\qquad i=1,2.
\eqno(24)$$
We  take the following  ansatz for  $\sigma$.
Let $m\in GL_p$ and let $q\in GL_s$, i.e.,
$m$ and $q$ are constant, invertible, respectively
$p\times p$ and  $s\times s$  matrices.
Define  the map
$\sigma_{m,q}: \MDS\rightarrow \MDS$ by
$$
\sigma_{m,q}: \pmatrix{\ell\cr z_+\cr z_-\cr w\cr}\mapsto
\pmatrix{ m\ell^\dagger m^{-1} \cr -m z_-^t q^{-1} \cr q z_+^t m^{-1}\cr
-q w^t q^{-1}\cr},
\eqno(25)$$
where $\ell^\dagger$ is given  by the standard
adjoint operation on  $PDO(p\times p)$,
$$
\ell^\dagger = (-1)^r\Delta^r \pa^r +\sum_{i=1}^r 
(-1)^{r-i} \pa^{r-i} u_i^t
\qquad\hbox{for}\qquad
\ell=\Delta^r \pa^r +\sum_{i=1}^r u_i \pa^{r-i}.
\eqno(26)$$
It is not hard to verify that  $\sigma_{m,q}$
satisfies (24)  whenever it maps the phase space
$\MDS$ to itself, which is ensured by the condition
$$
m \Delta^r m^{-1} = (-1)^r \Delta^r.
\eqno(27{\rm a})$$
The involutivity  of $\sigma_{m,q}$ leads to the further conditions
$$
m^t = \epsilon_m m,\quad \epsilon_m=\pm 1,
\qquad
q^t = \epsilon_q q,\quad \epsilon_q=\pm 1,
\quad\hbox{with}\quad \epsilon_m \epsilon_q=-1.
\eqno(27{\rm b})$$
Notice that if $\epsilon_m=-1$ then $p$ must be ${\it even}$ and
when $\epsilon_q=-1$ then $s$ must be even. 
For any natural numbers  $a$ and $b$
define  the $a\times a$ and $2b\times 2b$ matrices $\eta_a$ and
$\Omega_{2b}$ by
$$
\eta_a=\sum_{i=1}^a e_{i, a+1-i},
\qquad
\Omega_{2b}
=\sum_{i=1}^b e_{i, 2b +1-i} - \sum_{i=b+1}^{2b} e_{i, 2b+1-i},
\eqno(28{\rm a})$$
where the $e_{i,j}$ are elementary matrices of appropriate size
having a single nonzero entry $1$ at the $ij$ position. 
Let $\xi_{a}$ denote an  arbitrary $a\times a$ diagonal,
invertible matrix  subject to
$$
\eta_a \xi_{a} \eta_a = -\xi_{a},
\eqno(28{\rm b})$$
which means that $\xi_a$ is anti-symmetric under  transpose with
respect to the anti-diagonal.
Using this notation, 
we have the following types of  solutions for $\sigma_{m,q}$.
(The notion of a representative example  is justified later in this
section.)

\medskip\noindent
Type C1:   $r=2\rho$  even, $\forall\, p$,
$m$  is  diagonal and  $q$ is arbitrary with $\epsilon_q =-1$,
$s=2l$ even.
Representative example:  $\sigma_{\Delta, \Omega_{2l}}$.

\smallskip\noindent
Type C2:   $r=(2\rho +1)$  odd, $p=2k$  even,
$\Delta$ is such that
$\eta_p \Delta\eta_p=-\Delta$, $m=\xi_{p} \Omega_p$ and
$q$ is arbitrary with $\epsilon_q=-1$, $s=2l$ even.
Representative example: $\sigma_{\Delta \Omega_{2k}, \Omega_{2l}}$.

\smallskip\noindent
Type B: $r=(2\rho +1)$ odd, $p=2k$ even,
$\Delta$ is such that
$\eta_p \Delta\eta_p=-\Delta$,  $m=\xi_p \eta_p$ and
 $q$ is arbitrary with $\epsilon_q=+1$, $s=2l+1$ odd.
Representative example: $\sigma_{\Delta \eta_{2k}, \eta_{2l+1}}$.

\smallskip\noindent
Type D: $r=(2\rho +1)$ odd, $p=2k$ even,
$\Delta$ is such that
$\eta_p \Delta\eta_p=-\Delta$,  $m=\xi_p \eta_p$ and
 $q$ is arbitrary with $\epsilon_q=+1$,  $s=2l$ even.
Representative example: $\sigma_{\Delta \eta_{2k}, \eta_{2l}}$.

\medskip
Note that the condition $\eta_p \Delta\eta_p =-\Delta$,
which is present except for type C1, 
requires $\Delta$ to have the form
$\Delta={\rm diag}
\left(\Delta_1,\ldots,\Delta_k, -\Delta_k,\ldots, -\Delta_1\right)$,
where $p=2k$ and  $\Delta_i^r\neq \pm \Delta_j^r\neq 0$ for $i\neq j$ 
since $\Delta^r$ must have distinct, non-zero entries.
The Lie algebraic meaning of the notation 
referring to the various types will be explained below.

Given an involutive symmetry $\sigma=\sigma_{m,q}$, one finds that 
$\sigma :L\mapsto m L^\dagger m^{-1}$ for the Lax operator $L$ in (1).
It is not hard to see that this implies that the set of commuting 
Hamiltonians defined  by eq.~(8) 
admits a basis consisting of functions which are invariant or
anti-invariant (that change sign) with respect to the action of $\sigma$.
On account of (24), if $H\circ \sigma =H$ then the
Hamiltonian vector fields ${\bf X}_H^i$ are tangent to the
fixed point set $\MDS^\sigma \subset \MDS$ of $\sigma$.
Hence the  flows of a  ``discrete-reduced hierarchy''  may 
be defined by restricting the flows generated on $\MDS$
by the $\sigma$-invariant 
Hamiltonians in (8) to the fixed point set $\MDS^\sigma$.
These flows are bihamiltonian with respect to the restricted
Hamiltonians and a naturally induced bihamiltonian structure on
$\MDS^\sigma$.
The induced PBs on $\MDS^\sigma$ are defined by restricting the original 
PBs of functions of $\sigma$-invariant linear combinations of the 
components of $\ell, z_+, z_-, w$ --- which may be regarded as coordinates
on $\MDS^\sigma$ --- to $\MDS^\sigma$.
The Lax operator of the discrete-reduced system belongs to 
$$
M^\sigma=\pi(\MDS^\sigma)= \{ L\in M\,\vert\, L=mL^\dagger m^{-1}\,\}.
\eqno(29)$$ 

For fixed $p, r, s$ and a given symmetry
type C1, C2, B or D the various possible choices of $m$ and $q$
defining $\sigma_{m,q}$  are equivalent from the point of view of
the discrete reduction.  In fact, the fixed
point sets corresponding to two different  choices are always related
by a Poisson map of $\MDS$ given by
$$
(\ell, z_+, z_-, w)\mapsto
(\bar m\ell {\bar  m}^{-1}, \bar m z_+  {\bar q}^{-1},
{\bar q} z_-{\bar m}^{-1}, 
\bar q w {\bar q}^{-1})
\eqno(30)$$
with some constant matrices $\bar m\in GL_p$ and $\bar q\in GL_s$.
It is  in this sense that the examples we gave for the symmetries
of various types are {\it representative examples}.  

We mentioned that the  extended Gelfand-Dickey system follows from 
an application of the DS reduction procedure to the Lie algebra $gl_n$. 
As explained in particular cases in \q{\DF},
the  above discrete reductions are then 
induced by the reductions of $gl_n$ to a simple complex
Lie algebra $\G$ of $B$, $C$ or $D$ type.
This means that the discrete-reduced  systems
are associated with graded semisimple elements
of minimal positive grade from  certain  graded Heisenberg subalgebras
of $\G\otimes\Complex[\lambda,\lambda^{-1}]$ by means of
DS reduction (see also \q{\GHM,\McI} and the review in \q{\F}).
Since the graded Heisenberg
subalgebras of $\G\otimes\Complex[\lambda,\lambda^{-1}]$
are classified \q{\KP} by the conjugacy
classes \q{\Car} in the Weyl group ${\bf W}(\G)$ of $\G$,
we can label these generalised KdV hierarchies by the
respective conjugacy classes  in ${\bf W}(\G)$.
The conjugacy classes  that occur here
can be parametrised (as in \q{\Car,\DF})  by certain 
``signed partitions''.
The extended Gelfand-Dickey system itself
belongs to the conjugacy class $(r,\ldots, r, 1,\ldots, 1)$ 
of ${\bf W}(gl_n)$ given  by the partition in (2). 
Using this notation, we find that the above discrete symmetries  operate
on the generalised KdV  systems according to the following reduction rules:
$$\eqalign{
&\sigma_{\Delta, \Omega_{2l}}:
(\overbrace{ 2\rho, \ldots, 2\rho}^{p\rm\; times},
\overbrace{1,\ldots,1}^{2l\;\rm times})
\subset {\bf W}(gl_{2(p\rho +l)})
\quad \Longrightarrow \quad
(\overbrace{ \bar \rho,\ldots, \bar \rho}^{p\;\rm times},
\overbrace{1,\ldots,1}^{l\;\rm times})\subset
{\bf W}(C_{p\rho+l})\cr
&\sigma_{\Delta \Omega_{2k}, \Omega_{2l}}:
(\overbrace{ r, \ldots, r}^{2k\rm\; times},
\overbrace{1,\ldots,1}^{2l\;\rm times})
\subset {\bf W}(gl_{2(kr+l)})
\quad \Longrightarrow \quad
(\overbrace{  r,\ldots,  r}^{k\;\rm times},
\overbrace{1,\ldots,1}^{l\;\rm times}
)\subset  {\bf W}(C_{kr+l})\cr
&\sigma_{\Delta \eta_{2k}, \eta_{2l+1}}:
(\overbrace{ r, \ldots, r}^{2k\rm\; times},
\overbrace{1,\ldots,1}^{2l+1\;\rm times} )
\subset {\bf W}(gl_{2(kr+l)+1})
\  \Longrightarrow \
(\overbrace{  r,\ldots,  r}^{k\;\rm times},
\overbrace{1,\ldots,1}^{l\;\rm times}
)\subset
{\bf W}(B_{kr+l}) \cr
&\sigma_{\Delta \eta_{2k}, \eta_{2l}}:
(\overbrace{ r, \ldots, r}^{2k\rm\; times},
\overbrace{1,\ldots,1}^{2l\;\rm times})
\subset {\bf W}(gl_{2(kr+l)})
\quad \Longrightarrow \quad
(\overbrace{  r,\ldots,  r}^{k\;\rm times},
\overbrace{1,\ldots,1}^{l\;\rm times}
)\subset {\bf W}(D_{kr+l})\cr}
\eqno(31)$$
where $l\geq 0$ is arbitrary and $r=2\rho+1$ is odd.
With the aid of case by case  inspection,
this result has been established in \q{\DF} for $r>1$ and $l=0$.
Since the remaining cases  can be treated in a similar way,
we omit the proof (which simply amounts to diagram chasing).

One may try to lift the discrete symmetries given in the above
to analogous symmetries of the modified systems described in section 3.
Considering the modified systems that correspond to $L$ in (12),
one needs to find a lifted transformation rule
$\hat\sigma : \Theta \rightarrow \Theta$ for which 
$\mu\circ \hat \sigma = \sigma \circ \mu$, where 
$\mu: \Theta \rightarrow \MDS$ is the generalised Miura map.
It is clear that such a {\it local} map exists if and only if
the modification is symmetric in the sense that the same 
number of $(\pa +\theta)$ factors appears to the left and to the right 
of the special factor $K$ in (12).
This modification is available in the cases C2, B and D, 
for which $r=2\rho +1$ and  we have
$$
L=\Delta(\pa+\theta_1)\cdots
\Delta(\pa+\theta_{\rho})\Delta
\bigl[\pa+a-b\pad c\bigr]
\Delta(\pa+\theta_{\rho+1})\cdots\Delta(\pa+\theta_{2\rho})
\eqno(32)$$
by choosing $\kappa =\rho$ in (12).
The transformation rule
$\theta_i \mapsto \theta_i^{\hat \sigma}$ ($0,1,\ldots, 2\rho$)
is then not difficult to determine using  the requirement 
that  it must imply 
$L \mapsto  L^\sigma = m L^\dagger m^{-1}$  for $L$ in (32).
Of course $\hat\sigma$ is a Poisson map,
and the corresponding discrete-reduced hierarchy on the fixed point set
$\Theta^{\hat \sigma} \subset \Theta$ provides a modification 
of the hierarchy on $\MDS^\sigma\subset \MDS$. 
We leave it to the reader as an exercise to write down 
the explicit formula of $\hat\sigma$.

For the discrete symmetry of type C1 with $l>0$,  
a factorised Lax operator of the symmetric form is available only 
after performing 
the second factorisation of $K$ according to (15).
In this case  $r=2\rho$, and by choosing $\kappa = \rho-1$ in (12)
(and renaming the variables) we indeed obtain the
symmetric factorisation 
$$
L=\Delta(\pa+\theta_1)\cdots
\Delta(\pa+\theta_{\rho}) 
\bigl[{\bf 1}_p -\gamma (\pa+\vartheta +\beta \gamma )^{-1}\beta \bigr]
\Delta(\pa+\theta_{\rho+1})\cdots\Delta(\pa+\theta_{2\rho}).
\eqno(33)$$
The modified variables $\theta_i$ ($i=1,\ldots, 2\rho$)
and $\vartheta, \beta, \gamma$ now belong to the respective factors 
of the space 
$$
\Theta' = 
\bigl(\widetilde{gl}_p\bigr)^{2\rho}\times \widetilde{gl}_{2l}
\times \widetilde{\mat}(2l\times p)\times \widetilde{\mat}(p\times 2l).
\eqno(34)$$
The lifted action of the discrete symmetry on $\Theta'$ is easy to
determine explicitly using that for $L$ in (33) it must imply 
$L\mapsto L^\sigma$ with  $\sigma = \sigma_{\Delta, \Omega_{2l}}$.

\bigskip
\centerline{\bf 5.~Concluding remarks}
\medskip

We saw in section 4 that  many KdV type hierarchies 
that are associated with  certain conjugacy classes in the Weyl 
group ${\bf W}(\G)$ for $\G$ a classical  simple Lie algebra by 
generalised DS reduction \q{\GHM,\McI,\F} are also obtained as
discrete-reductions of extended matrix Gelfand-Dickey hierarchies. 
Note however that  not all KdV type  hierarchies 
based on a classical Lie algebra are discrete-reductions
of hierarchies  associated with  $gl_n$.
For example, a pseudodifferential operator model 
of the  KdV  system associated 
with the primitive regular  conjugacy class
$(\bar p,\bar p)$ in  ${\bf W}(D_{2p})$ by generalised DS
reduction is not known \q{\DF}.

In the DS approach  modifications 
of KdV type systems usually  correspond  to gauge transformations from
certain ``diagonal type gauges''  parametrised by the modified variables
to a  ``DS gauge''  parametrised by the KdV fields. 
The map $\mu:\Theta\rightarrow \MDS$ 
was obtained  in \q{\FM} by this method.
The modification $\nu': \Theta' \rightarrow \Theta$ 
mentioned after Proposition 4 also permits interpretation 
as a gauge transformation in the DS approach.
Moreover, the specific factorisations of $L$ in (32) and (33) 
that admit a local lifting of the relevant discrete  symmetry 
have a clear interpretation. 
Namely, these modifications correspond  to 
gauge sections that are mapped to themselves by the original 
discrete-symmetry transformation that operates on the first
order matrix differential operator variable used in the DS approach.
More details on the way discrete symmetries occur 
in the DS framework can be found in \q{\DF}.

\bigskip
\medskip
\noindent
{\bf Acknowledgement.}
The work of L.F.~was supported by the Alexander von Humboldt Foundation.

\vfill\eject

\bigskip
\centerline{\bf References}
\medskip

\bibitem{\Ch}
Yi Cheng, J.\ Math.\ Phys.\ {\bf 33} (1992) 3774.

\bibitem{\OS}
W.\ Oevel and W.\ Strampp,
Commun.\ Math.\ Phys.\ {\bf 157} (1993) 51.

\bibitem{\Oe}
W.\ Oevel, Physica {\bf A195} (1993) 533.

\bibitem{\De}
A.\ Deckmyn, Phys.\ Lett.\ {\bf B298} (1993) 318.

\bibitem{\Liu}
Q.P.\ Liu, Phys.\ Lett. {\bf A187} (1994) 373;
J.\ Math.\ Phys.\ {\bf 37} (1996) 2307.

\bibitem{\Che}
Yi Cheng, Commun.\ Math.\ Phys.\ {\bf 171} (1995) 661.

\bibitem{\Di}
L.A.~Dickey, Lett.\ Math.\ Phys.\ {\bf 34} (1995) 379.

\bibitem{\Arat}
H.\ Aratyn, L.A.\ Ferreira, J.F.\ Gomes and A.H.\ Zimerman,
preprint hep-th/9509096;
H.\ Aratyn, E.\ Nissimov and S.\ Pacheva, preprint hep-th/9701017.

\bibitem{\FM}
L.\ Feh\'er and I.\ Marshall,
Commun.\ Math.\ Phys.\ {\bf 183} (1997) 423.

\bibitem{\DS}
V.G.\ Drinfeld and V.V.\ Sokolov, J. Sov. Math. {\bf 30} (1985) 1975.

\bibitem{\KP}
V.G.\ Kac and D.H.\ Peterson,
pp.~276-298 in:~Proc.\ of Symposium on Anomalies, Geometry and Topology,
W.A.\ Bardeen and A.R.\ White (eds.),
World Scientific,  1985.

\bibitem{\GHM}
M.F.\ de Groot, T.J.\ Hollowood and  J.L.\ Miramontes,
Commun.\ Math.\ Phys.\ {\bf 145} (1992) 57.

\bibitem{\McI}
I.R.\ McIntosh,
{\it An Algebraic Study of Zero Curvature Equations},
PhD Thesis, Dept.\ Math., Imperial College, London, 1988 (unpublished).

\bibitem{\F}
L.\ Feh\'er,  talk given at the Marseille 1995 conference on 
${\cal W}$-algebras, hep-th/9510001.

\bibitem{\KW}
B.A.\ Kupershmidt and G.\ Wilson,  Invent.\ Math.\ {\bf 62} (1981) 403.

\bibitem{\FG}
A.P.\ Fordy and J.\ Gibbons, J.\ Math.\ Phys.\ {\bf 22} (1981) 1170.

\bibitem{\dBF}
J.\ de Boer and L.\ Feh\' er, Mod.\ Phys.\ Lett.\ {\bf A11} (1996) 317;
preprint hep-th/9611083.

\bibitem{\A}
M.\ Adler, Invent.\ Math.\ {\bf 50:3} (1979) 219.

\bibitem{\D}
L.A.\ Dickey, {\it Soliton Equations and Hamiltonian Systems},
Adv.\ Ser.\ Math.\ Phys., Vol.\ 12, World Scientific,   1991.

\bibitem{\DF}
F.\ Delduc and L.\ Feh\' er,
J.\ Phys.\ A: Math.\ Gen.\ {\bf 28} (1995) 5843.
                                                                           
\bibitem{\Car}
R.W.\ Carter, Comp.\ Math.\ {\bf 25} (1972) 1.

\bye